# Induced superconductivity and engineered Josephson tunneling devices in epitaxial (111)-oriented gold/vanadium heterostructures

Peng Wei[1,2*], Ferhat Katmis[1,2], Cui-Zu Chang[1], and Jagadeesh S. Moodera[1,2†]

1. Francis Bitter Magnet Laboratory, Massachusetts Institute of Technology, Cambridge, MA 02139, United States
2. Department of Physics, Massachusetts Institute of Technology, Cambridge, MA 02139, United States

*pwei@mit.edu; †moodera@mit.edu

**Abstract**: We report a unique experimental approach to create topological superconductors by inducing superconductivity into epitaxial metallic thin film with strong spin-orbit coupling. Utilizing molecular beam epitaxy technique under ultra-high vacuum condition, we are able to achieve (111) oriented single phase of gold (Au) thin film grown on a well-oriented vanadium (V) *s*-wave superconductor film with clean interface. We obtained atomically smooth Au thin films with thicknesses even down to below a nanometer showing near-ideal surface quality. The as-grown V/Au bilayer heterostructure exhibits superconducting transition at around 4 K. Clear Josephson tunneling and Andreev reflection are observed in S-I-S tunnel junctions fabricated from the epitaxial bi-layers. The barrier thickness dependent tunneling and the associated subharmonic gap structures (SGS) confirmed the induced superconductivity in Au (111), paving the way for engineering thin film heterostructure based *p*-wave superconductors and nano devices for Majorana fermion.

The quasiparticles at an interface of thin film heterostructures can lead to the observation of emerging properties. It has been shown theoretically that epitaxial noble metal (heavy element) thin films with well-defined crystalline surface, such as Ag(111) and Au(111) etc., are ideal materials for robust $p + ip$ superconductors due to giant Rashba spin-orbit coupling (SOC).[1-3] The surface quasiparticles gain this SOC owing to the nature of broken inversion symmetry and the short Fermi wave length of the metallic surface states.[2] As a result, the magnitude of SOC is several orders larger than that in many semiconductor nanowires.[2,4-10] When the $s$-wave superconductivity (SC) is induced into these epitaxial heavy metal thin films, topological superconductivity and Majorana fermions is expected to emerge in the presence of Zeeman splitting (ZS) that can be generated either by an external magnetic field or interface exchange field.[11,12] From experimental point of view, creating hybridized quasiparticle with combined interactions of SOC, ZS and SC is achievable via interfacial proximity effect in hybrid thin film layers.[11,13] However, the epitaxial growth of ultra-thin noble metal films with well-ordered crystalline surface on a superconductor still remains challenging.[14] Due to the large surface energy of Ag or Au the grown layer usually forms island type growth mode, which prevents them from forming continuous and smooth films in thicknesses down to a few nanometers.[15] Magnetic seed layers, such as Fe, Co etc., have been reported as buffer for growing thin (< 50 nm) Ag or Au films,[16,17] whereas it does not allow inducing superconductivity. Non-magnetic seed layer (Nb) has been demonstrated to facilitate the growth of Au thin film;[18] however, the Nb is too thin (~ 1 nm) to show any superconductivity.[14]

Here we demonstrate high quality ultra-thin (~ 4 nm) (111)-oriented Au films grown on $s$-wave superconductor – vanadium (V) that exhibits clear superconducting transition at $T_c$ ~ 4 K. The in-situ reflection high-energy electron diffraction (RHEED) studies and ex-situ low-angle X-ray reflectivity (XRR) and X-ray diffraction (XRD) studies confirm high quality (111)-oriented surface of gold that shows high quality even with thickness down to 0.4nm. The induced SC in Au is confirmed by tunneling using vertically fabricated Josephson junctions, in which the superconducting tunneling is further tuned by varying the tunnel barrier thickness within ±1Å.

The epitaxial growth of Au on V involved a refined multi-step growth sequence as demonstrated in Fig. 1 with RHEED snapshots ( see also Methods). The RHEED studies were performed at different critical stages while growing. Due to glancing angle of incidence, the electron beam has short penetration depth. Therefore it is sensitive to the crystallinity of the thin film surface. Besides, the azimuthal angle between the electron beam and the substrate can be adjusted (Fig. 2a) to probe different crystalline axes of the surface. Initially we direct the electron beam along the $[2\bar{1}\bar{1}0]$ axis of the (0001) surface of $Al_2O_3$. This RHEED pattern is shown at $t = 0$ in Fig. 1. The clearly observed higher-order Laue rings confirm the high quality of (0001)-$Al_2O_3$ surface. Subsequently, a 1 nm seed layer of V was grown at room temperature, which forms a complete diffusive image by blanking the substrate diffraction spots ($t = t_1$, Fig. 1). The vanishing RHEED pattern is an indication of complete wetting of the sapphire surface by formation of an amorphous V seed layer. This V passivation layer serves to improve the growth quality of the subsequent layers. To ensure the high crystal quality of the subsequently grown 20 nm V layer, we annealed the seed layer by raising the substrate temperature gradually to 490 °C in ~ 1 hour. During this process, the RHEED pattern was continuously monitored. At $t = t_2$, sharp diffraction streaks appeared that revealed the crystallization of the V seed layer (Fig. 1). Then the substrate temperature was quickly reduced to ~ 460 °C followed by the growth of 20 nm V layer at a rate of ~ 0.1 Å/sec. As the V layer gets thicker, the streak-like RHEED pattern becomes spot-like (Fig. 1 at $t = t_3$), indicating the transition into three-dimensional (3-D) growth mode instead of the two-dimensional (2-D) growth mode.[19] As soon as the signature of RHEED spots was observed, the growth of V was terminated. Then Au was quickly deposited over it (rate 0.03 Å/sec) at a substrate temperature ~ 410 °C (Fig. 1 at $t = t_4$). The Au layer showed 2-D growth mode as seen by the reappearance of streaky RHEED pattern even with a coverage of 0.4nm thick layer (close to 2 monolayers) (Fig. 1 at $t = t_4$). This observation indicates the growth of high quality ultrathin gold films on vanadium. As Au gets thicker (approaching 4nm), the RHEED streak pattern became narrower and sharper (Fig. 1 at $t = t_5$) that further demonstrates the enhanced surface quality.

The terminating surface of the 4nm Au was found to be (111)-oriented. This was confirmed by studying the RHEED pattern while varying the azimuthal angle of the incident electron beam. Fig. 2a illustrates the corresponding experimental setup. Fig. 2b and 2c show two characteristic RHEED patterns that periodically appear when the azimuthal angle is changed by 60°. This six-fold rotation symmetry is a direct proof of the (111)-orientation of the grown Au surface. Further epitaxial relationship between the Au layer and the substrate was investigated by first aligning the electron beam along the $[2\bar{1}\bar{1}0]$ and the $[1\bar{1}00]$ axes of (0001)-Al$_2$O$_3$, then monitoring the RHEED during Au growth. We note here that these two crystalline axes have the azimuthal angel difference of 30°. While sharp streaky RHEED was seen along both directions, the steaks have two distinct spacing (Fig. 2b and 2c), which is the measure of the atomic line (rods) spacing of a two-dimensional lattice in the reciprocal space.[19] In fact, the spacing between the RHEED streaks $W$ and the spacing between atomic lines (rods) of the Au surface $a$, are connected by: $a = \dfrac{\lambda L}{W}$, where $L$ is the distance between the RHEED screen and the sample, $\lambda$ is the wave length of the incident electron beam. Since an electron beam with a high acceleration energy (15 keV) was used, the relativistic correction to the electron's wave length $\lambda$ gives $\lambda = 12.3/[V(1+1.95 \times 10^{-6} \cdot V)]^{1/2}$ Å.[19] With $V$ = 15 kV, we have $\lambda \sim 0.099$ Å. The distance between our RHEED screen and the sample is $L$ = 32.13 ± 0.50 cm with the error accounting for the location of the electron beam spot on the sample substrate. The spacing of the RHEED streaks was measured to be 12.78 ± 0.50 mm (Fig. 2b) and 21.78 ± 0.60 mm (Fig. 2c), where the errors represent half width of the RHEED streak lines. As a result, the estimated atomic line (rod) spacing is $a_1$ = 2.48 ± 0.10 Å (Fig. 2b) and $a_2$ = 1.46 ± 0.05 Å (Fig. 2c). It is clear that $a_1 \approx \sqrt{3} a_2$, which points to two distinct atomic line (rod) directions on the (111) surface. Fig. 2d shows the schematics of these two atomic line (rod) directions, where $a_1$ is the spacing between the atomic lines along [1$\bar{1}$0] direction of the f.c.c. lattice and $a_2$ is the spacing between lines along [$\bar{1}\bar{1}$2] direction. Using $a_1$ and $a_2$, we further estimate the lattice constant of Au as $a = \sqrt{2} \cdot \sqrt{(a_1^2 + a_2^2)}$ = 4.07 ± 0.25 Å, which matches well with other reported values.[16] In fact, these in-situ RHEED results tell us the epitaxial relation between Au and the substrate as: Au [1$\bar{1}$0] // Al$_2$O$_3$[2$\bar{1}\bar{1}$0].

The crystallinity and surface/interface roughness of the thin film bilayer are further elucidated by the low angle XRR (Fig. 3a) and high angle XRD $\Theta/2\Theta$ (Fig. 3b) scans. In Fig. 3a, the XRR demonstrates clear oscillations that are extended to the high angle $2\Theta$ >10 degree, which reveals the smooth surface/interface of the bilayer heterostructure.[20] Quantitative fittings were performed (red line in Fig. 3a) which matched well with the XRR oscillations (symbols in Fig. 3a). The XRD data shows major Bragg peak at the $2\Theta$ angle of the (111) reflection of Au(111) indicating high crystallinity quality. Remarkably, sharp Laue oscillations also occur near the Bragg peaks of Au and V, which again indicates sharp surface/interface of the heterostructure. These Laue oscillations clearly have two different periods and we note them as $\Delta 2\Theta_{Au}$ and $\Delta 2\Theta_V$ (Fig. 3b). From these two distinct periodicities, we calculate the thickness of Au as ~ 4.7 nm and V as ~ 21.2 nm, which matches quite well to the thicknesses monitored by the quartz crystal sensor during the growth. We note here that the Bragg peak of V(110) (Fig. 3b) points to the (110) oriented 20nm V film. While the 1 nm V seed layer shows (111) orientation from the in-situ RHEED measurements, the enhanced RHEED spotty features for 20nm V indicate the change of growth feature when V layer becomes thicker (Fig. 1 $t = t_2$ vs. $t = t_3$). Since bulk V has a body centered cubic (bcc) structure, thick film of V tends to grow along (110) direction.

The growth conditions can be well corroborated by the surface morphology studies, by the SEM studies performed immediately after the growth. We found that the growth temperatures of the Au and V layers critically affected the morphology. By fine tuning the growth conditions, we were able to observe the transitions from island-like growth to island-less growth. Fig. 3c – 3e show the SEM images of a set of three heterostructure thin film samples, where the scale bar denotes 400 nm. All of them have 1 nm V seed layer grown at room temperature. The sample in Fig. 3c has V grown at 634 °C and Au grown at 318 °C. Clear holes and patches can be seen suggesting island-like growth. Increasing the growth temperature of Au to 410 °C while keeping the growth temperature for V same removed the holes, whereas the patches can still be seen (Fig. 3d). The best quality samples were achieved by reducing the growth temperature of V to 450 °C and keeping the growth of Au at ~ 410 °C, in which case the SEM shows island-less morphology of Au on V (Fig. 3e).

The superconductivity (SC) of the bilayer heterostructures was studied by Cooper pair tunneling at temperature $T = 1$ K. We fabricated a layer of $Al_2O_3$ tunnel barrier and a top superconducting electrode Al over the Au (111) surface to form S-I-S tunnel junctions (Fig. 4a). The tunneling takes place between Au and Al electrodes through $Al_2O_3$ barrier. By carefully varying the thickness of the $Al_2O_3$ by ± 0.1 nm around 1.5 nm, we were further able to produce a systematic change of the junction resistances $R_{Jn}$, which in turn modulates the SC tunneling behavior. Here we note that the set of tunnel junctions were fabricated on the same V/Au bilayer stripe and $R_{Jn}$ was measured in a four-terminal manner across the $Al_2O_3$ barrier (Fig. 4a). Hence the change of $R_{Jn}$ reflects the change in the barrier thickness.

In low resistance junction, $R_{Jn} \sim 10\ \Omega$, Josephson supercurrent was observed (Fig. 4b). This can be attributed to Cooper pair tunneling across ultrathin $Al_2O_3$ barrier in between Al and Au (111), where SC is induced into Au (111) by V. To further elucidate it, we progressively increased the $R_{Jn}$ of the tunnel junctions from $R_{Jn} \sim 0.6\ k\Omega$ to $4.8\ k\Omega$ (Fig. 4c), where multiple Andreev reflection was observed. We found that the Andreev peak at $V_{bias} = 0$ mV sharply decreased as the $R_{Jn}$ (or $Al_2O_3$ barrier thickness) increased (Fig. 4c). Since the Andreev peak at $V_{bias} = 0$ mV is a direct signature of Cooper pair tunneling, its sharp decrease upon increasing $R_{Jn}$ (or $Al_2O_3$ barrier thickness) points to the fact that tunneling occurs across the $Al_2O_3$ layer. If one estimates the junction resistance for 1.5 nm thick $Al_2O_3$ barrier by changing its thickness by ± 0.1 nm, the junction resistance changes by almost an order of magnitude,[21] in very good agreement with what is seen here. Thus it supports the induced SC in Au (111) due to the superconducting V underneath. Furthermore, the subharmonic gap structures (SGS) of the Andreev reflection, i.e. the sharp drop of the $dI/dV$ is at similar $V_{bias}$ in different samples regardless of $R_{Jn}$ (Fig. 4d). The tunneling conductance $dI/dV$ follows the effective number of the Andreev reflections. In asymmetric S-I-S' junction as in our case, the SGS takes place when there is a loss of the Andreev reflection, seen by the dips in conductance which are located for example at $eV_{bias} = \Delta_1$, $\varDelta_2$ and $\varDelta_2$ - $\varDelta_1$ where $\varDelta_1$ and $\varDelta_2$ are the SC gaps ($\varDelta_2 > \varDelta_1$).[22] Therefore, the SGS positions reflect the gaps of the superconductors. From Fig. 4d (SGS marked by dashed lines), we can get $\varDelta_1 \sim$ 0.24 meV, which matches very well to the SC gap of 14 nm evaporated Al thin film.[23,24] Since the Cooper pair tunneling is proven to take place between Al and Au, the second gap $\varDelta_2 \sim 0.6$ meV is attributed to the induced SC gap in Au(111) due to proximity effect. The other $dI/dV$ dip feature that satisfies $eV_{bias} = \varDelta_2 - \varDelta_1$ is also observable that is indicated by the arrow in Fig. 4d. The SGS further supports the induced SC in gold, where a gap as large as 0.6 meV is obtained that matches well to the $T_c$ in V/Au measured by the resistance drop of the bi-layer stripe. The successfully achieved high quality (111)-oriented superconducting Au thin films, where giant spin-orbit coupling exists, demonstrated an ideal thin film heterostructure system for topological superconductivity and for seeking Majorana fermions. Our findings pave the way for investigating unknown SC in hybrid systems in the presence of various material interactions, and can lead to the observations of new phenomena in physics.[2,25]

**Figures and captions:**

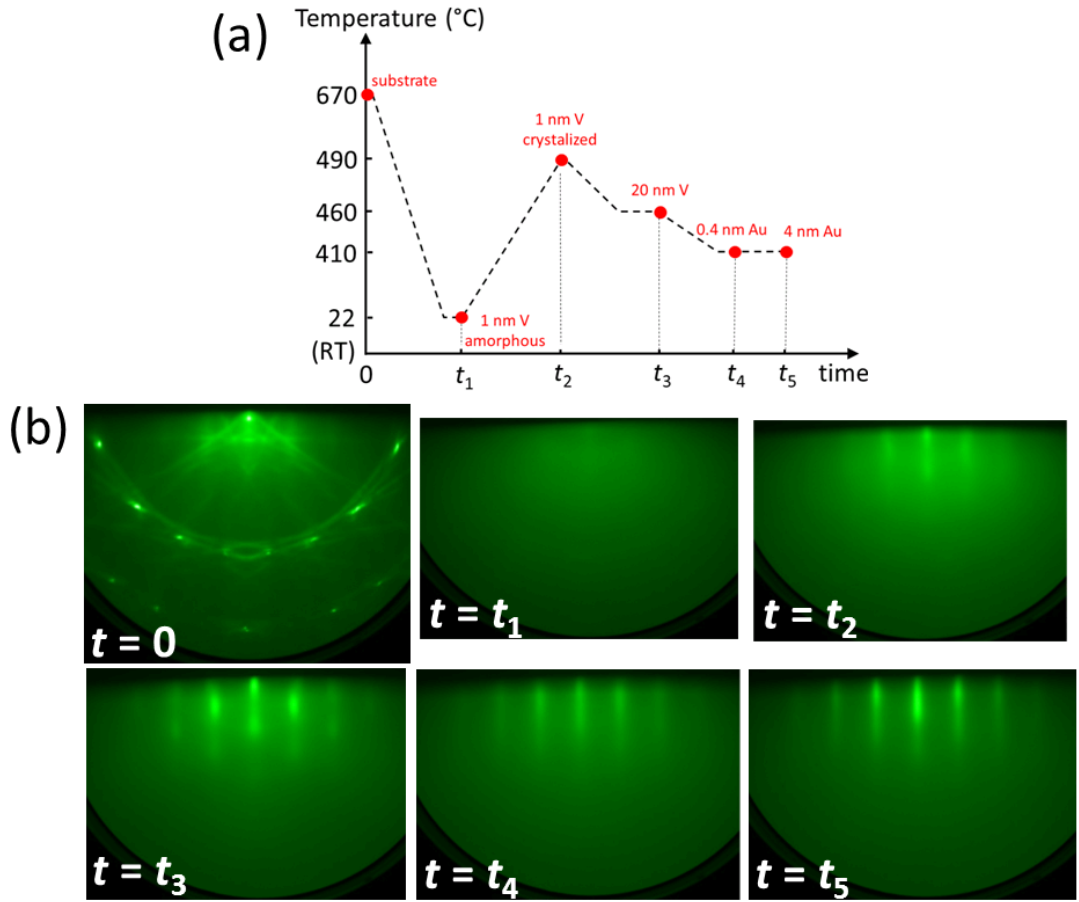

**Fig. 1. (a)** Detailed growth sequence for the thin film heterostructure of V/Au. At $t = 0$, the sapphire substrate is processed and ready for the growth. At $t = t_1$, 1 nm V seed layer is grown at room temperature. At $t = t_2$, the V seed layer starts to crystallize upon annealing. At $t = t_3$, 20 nm V layer is grown at ~ 460 °C. At $t = t_4$, 0.4 nm (~ 2 monolayers) Au is grown at ~ 410 °C. At $t = t_5$, 4 nm Au layer is grown. **(b)** The corresponding RHEED patterns at each specified time. Note here the well-defined RHEED streaks for the 0.4 nm Au grown on V ($t = t_4$), and the sharper streaks for the 4 nm Au grown on V ($t = t_5$).

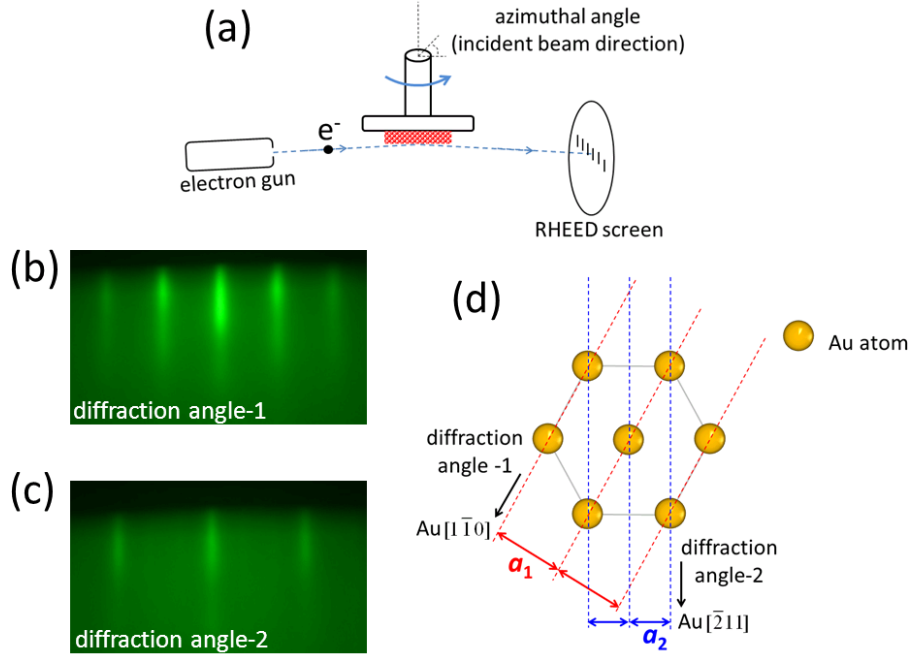

**Fig. 2. (a)** The schematics of the RHEED set up with variable electron incidence angle (azimuthal angle). This method is utilized to pin point the rotation symmetry and crystal orientation of the thin film. **(b)** The RHEED pattern of 4 nm gold on V along the $[1\bar{1}0]$ direction of its (111) surface. **(c)** Upon rotating the azimuthal angle by 30 °, as shown in **(a)**, the spacing of the RHEED streaks increases, which corresponds to the $[\bar{2}11]$ direction on the (111) surface of Au. The two distinct RHEED patterns repeat themselves when the azimuthal angle rotates every 30 °. **(d)** Schematic showing the method to extract the periodic atomic lines (rods) spacing, marked by red and blue dashed lines, thereby extracting the lattice constant of the thin film and further verify its crystal orientation.

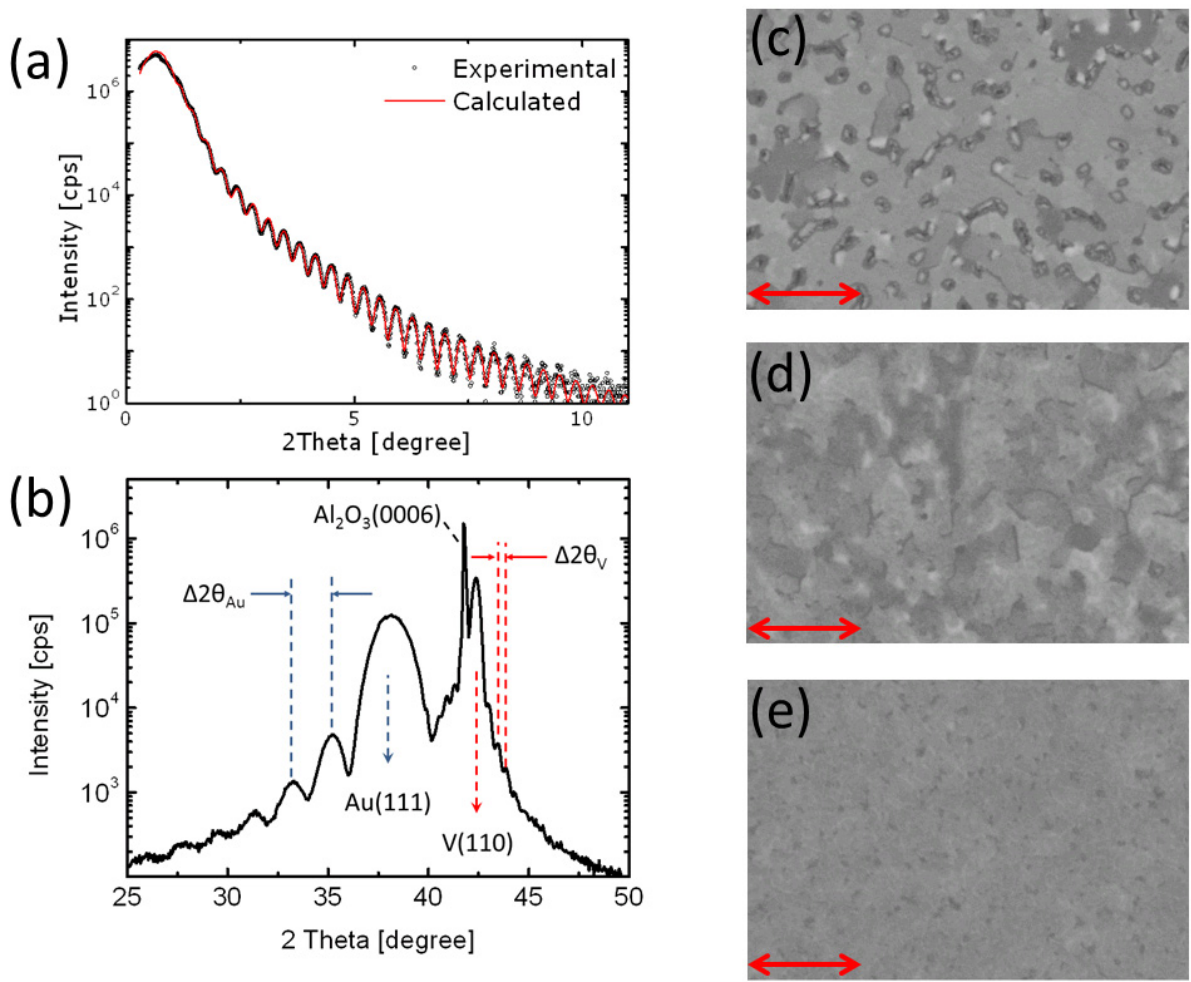

**Fig. 3. (a)** The representative data of ex-situ low-angle X-ray reflectivity (XRR) on V/Au bilayer samples. The well-defined oscillations are due to the interference of the X-ray reflected from material surface/interface. These oscillations persist until large incidence angle, which proves the smooth film surface/interface. **(b)** The representative data of ex-situ X-ray diffraction (XRD) studies. Besides the characteristic Au(111) and V(110) peaks, pronounced Laue oscillations are seen. The periodicity $\Delta 2\Theta_{Au}$ and $\Delta 2\Theta_V$ directly reflects the thicknesses separately for the Au and V layer. Observing clear Laue oscillations in addition to the characteristic Bragg peaks further confirms the high quality V/Au heterostructure interface. **(c) – (e)** The surface morphology of Au thin film on V layer as examined by SEM. The scale bar represents 400 nm. The comparison is for three samples under identical growth conditions except that: **(a)** 20 nm V grown at 634 °C and Au grown at 318 °C - pin holes are clearly seen; **(b)** 20 nm V grown at 634 °C and Au grown at 410 °C - pin holes are less and grains start to merge whereas patches (islands) still exist. **(c)** 20 nm V grown at 450 °C and Au grown at 410 °C - best quality smooth Au film.

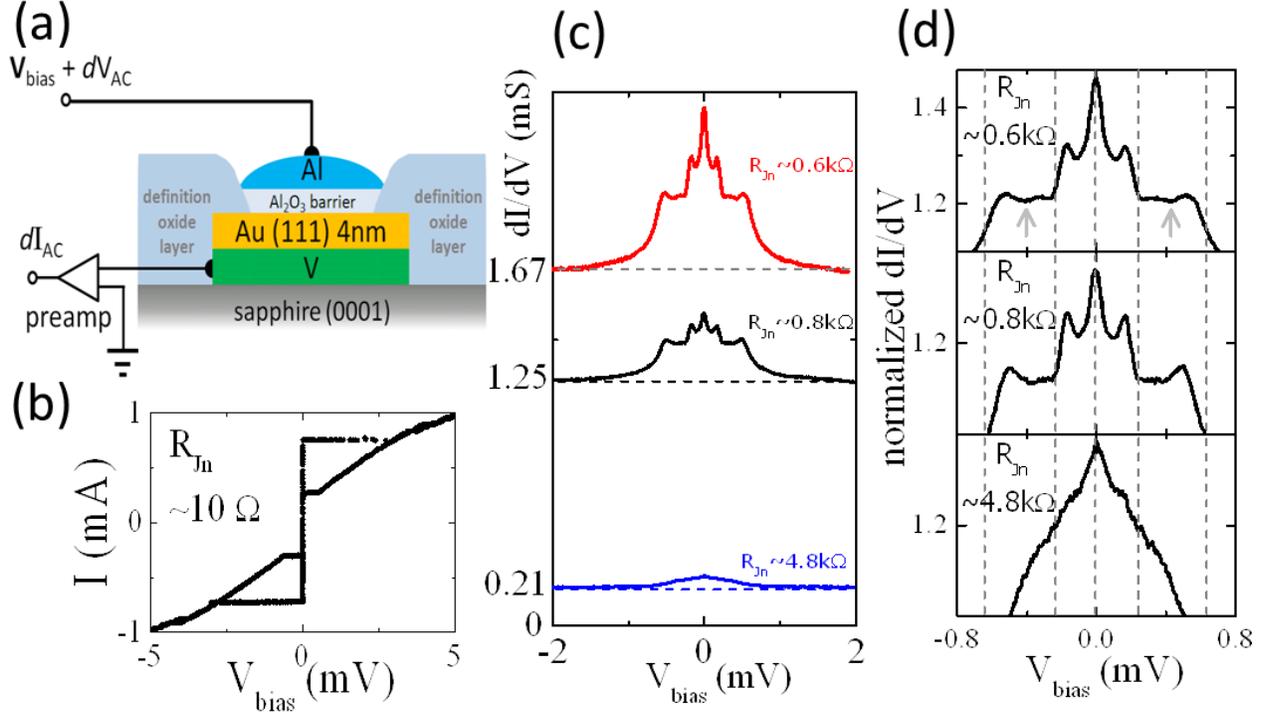

**Fig. 4. (a)** The schematic of S-I-S tunnel junction fabricated from the epitaxial V/Au bilayer samples. **(b)** In the lowest resistance junction ($R_{Jn} \sim 10\ \Omega$), clear Josephson supercurrent is observed. The junction resistance is measured by four-terminal method. Note here the high $I_C R_N$ value ~ 7.3 mV. **(c)** Systematically tuning of the junction resistance by merely increasing the thickness of the $Al_2O_3$ barrier. Three samples are made simultaneously on the same V/Au bilayer stripe only by varying the $Al_2O_3$ thickness, which results in junction resistance $R_{Jn} \sim$ 0.6 k$\Omega$, 0.8 k$\Omega$, and 4.8 k$\Omega$ respectively. It is clearly shown that the Andreev reflection peak at $V_{bias}$ = 0 mV decreases when the junction resistance, or the $Al_2O_3$ barrier thickness, increases. This observation directly supports the fact that the Andreev reflection takes place between Al and Au (across $Al_2O_3$). Therefore, Au is superconducting. **(d)** The peak (energy) positions of the SGS are seen to be independent of the junction resistance. The SGS takes place at $eV_{bias} = \Delta_1$, $\Delta_2$ and $\Delta_2 - \Delta_1$. Therefore, it gives the gap size of the two superconductors: Al gap $\Delta_1 \sim 0.24$ meV and the induced SC gap in Au is $\Delta_2 \sim 0.6$ meV. Both of them match well with the measured $T_c$ of the superconductors.

**Methods:**

The growth of the V/Au bi-layer heterostructure was carried out in a custom built molecular beam epitaxy (MBE) chamber under ultra-high vacuum (UHV) environment (pressure $\sim$ low $10^{-10}$ torr). High-purity (99.995%) V and Au sources were melted in a multi-pocket electron beam evaporator. During the deposition, the film thickness and growth rate were monitored by a quartz crystal sensor. The quality of the bi-layers and their interfaces were examined in-situ with RHEED, which has 15 keV electron beam directed at glancing angle ($\sim 2$ degree) to the thin film surface. The chosen substrates were 0.5 mm polished single crystal $\alpha$-$Al_2O_3$(0001) (sapphire). Prior to the growth, the substrates were sonicated in acetone and IPA followed by standard SC1 cleaning. They were further processed in the MBE chamber by vacuum annealing at $\sim 670$ °C for about an hour until the sharp diffraction patterns were observed by RHEED confirming high quality substrate surface. The post ex-situ high-resolution XRR and XRD were carried out using 1.54 Å Cu-$K_{\alpha 1}$ radiation. The multilayer Josephson tunnel junctions were fabricated in a separate high vacuum chamber using in-situ manipulated shadow masks for the control of different tunnel barrier thicknesses. The Josephson tunneling and Andreev reflection characteristics were measured in a Faraday cage and LHe bath cryostat at a temperature $\sim$ 1K.

**Acknowledgments:**

The authors acknowledge the enlightened discussions with P. A. Lee and A. C. Potter. P.W. and J.S.M. would like to acknowledge the support from the John Templeton Foundation Grant No. 39944. P.W., F.K., C.-Z.C. and J.S.M. would like to acknowledge the support from National Science Foundation Grant DMR-1207469, and Office of Naval Research Grant N00014-13-1-0301.

**References:**

1  Potter, A. C. & Lee, P. A., Multichannel Generalization of Kitaev's Majorana End States and a Practical Route to Realize Them in Thin Films, *Phys Rev Lett* **105**, 227003, (2010).

2  Potter, A. C. & Lee, P. A., Topological superconductivity and Majorana fermions in metallic surface states, *Phys Rev B* **85**, 094516, (2012).

3  Liu, J., Potter, A. C., Law, K. T. & Lee, P. A., Zero-Bias Peaks in the Tunneling Conductance of Spin-Orbit-Coupled Superconducting Wires with and without Majorana End-States, *Phys Rev Lett* **109**, 267002, (2012).

4  Nicolay, G., Reinert, F., uuml, fner, S. & Blaha, P., Spin-orbit splitting of the L-gap surface state on Au(111) and Ag(111), *Phys Rev B* **65**, 033407, (2001).

5  Mourik, V. *et al.*, Signatures of Majorana Fermions in Hybrid Superconductor-Semiconductor Nanowire Devices, *Science* **336**, 1003-1007, (2012).

6  Das, A. *et al.*, Zero-bias peaks and splitting in an Al-InAs nanowire topological superconductor as a signature of Majorana fermions, *Nat Phys* **8**, 887-895, (2012).

7  Deng, M. T. *et al.*, Anomalous Zero-Bias Conductance Peak in a Nb–InSb Nanowire–Nb Hybrid Device, *Nano Lett* **12**, 6414-6419, (2012).

8  Finck, A. D. K., Van Harlingen, D. J., Mohseni, P. K., Jung, K. & Li, X., Anomalous Modulation of a Zero-Bias Peak in a Hybrid Nanowire-Superconductor Device, *Phys Rev Lett* **110**, 126406, (2013).

9  Lee, E. J. H. *et al.*, Spin-resolved Andreev levels and parity crossings in hybrid superconductor-semiconductor nanostructures, *Nat Nano* **9**, 79-84, (2014).

10 Larsen, T. W. *et al.*, Semiconductor-Nanowire-Based Superconducting Qubit, *Phys Rev Lett* **115**, 127001, (2015).

11 Wei, P. *et al.*, Exchange-Coupling-Induced Symmetry Breaking in Topological Insulators, *Phys Rev Lett* **110**, 186807, (2013).

12 Li, B. *et al.*, Superconducting Spin Switch with Infinite Magnetoresistance Induced by an Internal Exchange Field, *Phys Rev Lett* **110**, 097001, (2013).

13  Wei, P. *et al.*, Strong Interfacial Exchange Field in 2D Material/Magnetic-Insulator Heterostructures: Graphene/EuS, *arXiv:1510.05920*, (2015).

14  Kästle, G., Boyen, H. G., Schröder, A., Plettl, A. & Ziemann, P., Size effect of the resistivity of thin epitaxial gold films, *Phys Rev B* **70**, 165414, (2004).

15  Vook, R. & Oral, B., The epitaxy of gold, *Gold Bull* **20**, 13-20, (1987).

16  Kamiko, M. & Yamamoto, R., Epitaxial growth of Au(1,1,1) on α-$Al_2O_3$(0,0,0,1) by using a Co seed layer, *J Cryst Growth* **293**, 216-222, (2006).

17  Paggel, J. J., Miller, T. & Chiang, T. C., Quasiparticle Lifetime in Macroscopically Uniform Ag/Fe(100) Quantum Wells, *Phys Rev Lett* **81**, 5632-5635, (1998).

18  Kästle, G. *et al.*, Growth of thin, flat, epitaxial (1,1,1) oriented gold films on c-cut sapphire, *Surf Sci* **498**, 168-174, (2002).

19  Mahan, J. E., Geib, K. M., Robinson, G. Y. & Long, R. G., A review of the geometrical fundamentals of reflection high‐energy electron diffraction with application to silicon surfaces, *Journal of Vacuum Science & Technology A* **8**, 3692-3700, (1990).

20  Daillant, J. & Gibaud, A., X-ray and Neutron Reflectivity, *Book (Lecture Notes in Physics), Springer* **Volume 770 2009**, (2009).

21  Dorneles, L. S., Schaefer, D. M., Carara, M. & Schelp, L. F., The use of Simmons' equation to quantify the insulating barrier parameters in Al/AlOx/Al tunnel junctions, *Appl Phys Lett* **82**, 2832-2834, (2003).

22  Hurd, M., Datta, S. & Bagwell, P. F., Current-voltage relation for asymmetric ballistic superconducting junctions, *Phys Rev B* **54**, 6557-6567, (1996).

23  Meservey, R. & Tedrow, P. M., Properties of Very Thin Aluminum Films, *J Appl Phys* **42**, 51-53, (1971).

24  Tedrow, P. M. & Meservey, R., Spin-Paramagnetic Effects in Superconducting Aluminum Films, *Phys Rev B* **8**, 5098-5108, (1973).

25  Manchon, A., Koo, H. C., Nitta, J., Frolov, S. M. & Duine, R. A., New perspectives for Rashba spin-orbit coupling, *Nat Mater* **14**, 871-882, (2015).